\documentclass[preprint]{ptephy_v1}

\preprintnumber{XXXX-XXXX} 
\usepackage{hyperref}

\begin{document}

\title{Real-time alpha-ray track imaging using a synthetic diamond scintillator}


\author{Atsuhiro Umemoto}
\affil{International Center for Quantum-field Measurement Systems for Studies of the Universe and Particles (QUP),
High Energy Accelerator Research Organization (KEK), 1-1 Oho, Tsukuba, Ibaraki 305-0801,Japan
\email{aumemoto@post.kek.jp}}

\author[2]{Masao Yoshino}
\affil[2]{Institute for Material Research, Tohoku University, Miyagi 980-8577, Japan}

\author[3]{Takashi Iida}
\affil[3]{Faculty of Pure and Applied Sciences, University of Tsukuba,  Tennoudai, Tsukuba, Ibaraki, 305-8571, Japan}

\author[4]{Masashi Miyakawa}
\author[4]{Takashi Taniguchi}
\affil[4]{Research Center for Materials Nanoarchitectonics, National Institute for Materials Science, 1-1 Namiki, Tsukuba, Ibaraki 305-0044, Japan}

\author[5]{Shintaro Nomura}
\affil[5]{Division of Physics, University of Tsukuba, Tennoudai, Tsukuba, Ibaraki, 305-8571, Japan}


\begin{abstract}%
We demonstrated real-time imaging of alpha-ray tracks using a high-pressure, high-temperature synthetic diamond scintillator. The optical microscopy system consisted of a magnification objective lens and an electron-multiplying charge-coupled device (EMCCD) to detect luminescence from the diamond substrate irradiated with alpha-rays from a $^{241}$Am source. Optical images of alpha-ray tracks were successfully captured under ambient conditions, and image processing methods were performed to enable automatic event recognition and track length detection. These results demonstrate that diamond is a promising scintillator material for alpha-ray imaging, representing an important step toward exploiting the potential of diamond scintillators.

\end{abstract}

\subjectindex{xxxx, xxx}

\maketitle

\section{Introduction}
A scintillator plate enables real-time imaging of charged particle tracks when combined with a microscopy system, providing not only event-rate and energy-spectrum measurements but also spatial localization of the reaction point.
This method has been primarily developed for detecting alpha-ray tracks, as it is not sensitive to beta-rays due to the difference in energy deposition.  The alpha imagers have been developed for specific applications, such as radionuclide detection in cellular environments for targeted alpha-particle therapy~\cite{TAPT_review}~\cite{TAPT_Yamamoto}, and evaluation of internal exposure dose at nuclear facilities~\cite{alpha_kurosawa}. 
Not all scintillators can function as alpha imagers; for example, stability at room temperature under atmospheric pressure (i.e., non-hygroscopic properties) and high light yield are required.
In previous studies, Gd$_{3}$Al$_{2}$Ga$_{3}$O$_{12}$(GAGG)~\cite{alpha_GAGG} and ZnS(Ag)~\cite{alpha_ZnS} are used as scintillator plates.

Diamond has emerged as a promising candidate material for use as a scintillator plate in alpha imaging applications. 
Its chemical stability and mechanical robustness are well recognized, and impurity-related defects introduced into its wide bandgap of 5.47 eV can function as color centers in the visible wavelength range. 
In addition,  the radiation hardness of diamond is a critical advantage for ensuring stable operation in high-radiation environments.
Our recent investigations have demonstrated that nitrogen-containing diamond synthesized by the high-pressure, high-temperature (HPHT)~\cite{HPHT} method can achieve a high light yield of (5.5$\pm$0.5) $\times$ 10$^{4}$ photons/MeV~\cite{DiamondScintillator1}.
Building upon these results, the present study focuses on the synthesis and comprehensive characterization of nitrogen-doped HPHT diamond, with the objective of advancing its practical implementation as a scintillator detector.

In this paper, we demonstrated real-time alpha-ray track imaging using our synthetic HPHT diamond.
A custom-built microscopy system, comprising a magnification objective lens and an electron-multiplying charge-coupled device (EMCCD), was employed to capture images of alpha-ray tracks via the luminescence emitted from the diamond.
Image processing was performed to enable automatic event recognition and extract track information. These results confirm the viability of diamond as a scintillator material for alpha-ray imaging and highlight its potential for use in high-resolution, real-time radiation imaging systems.

\section{Methods}
\subsection{Diamond substrate}
\begin{figure}[htbp]
 \begin{center}
 \includegraphics[width=14cm]{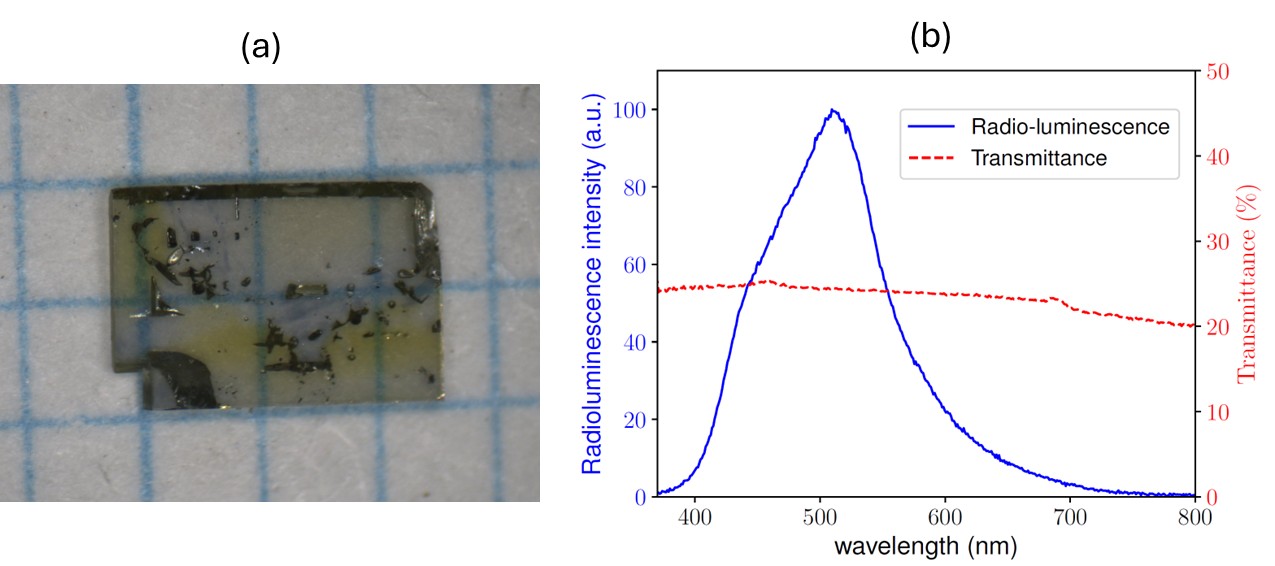}
\caption{Photographic and spectroscopic characterization of the diamond substrate used in this study: (a) photograph taken on 1 mm graph paper, and (b) X-ray excited radioluminescence and optical transmittance spectra.}
\label{fig:DiamondPhoto}
 \end{center}
\end{figure}

The diamond crystal was synthesized using a modified belt-type HPHT apparatus with a cylinder bore diameter of 32 mm at the National Institute for Materials Science~\cite{HPHT1}.
A single-crystal of diamond was grown on a (100) plane seed crystal in a cobalt solvent (Co-0.4 wt$\%$ B; Rare Metallic Co., Ltd.) using the temperature gradient method at 5.5 GPa and a temperature of 1430-1480$^{\circ}$C for 55 hours~\cite{HPHT2}.
High-purity graphite (Tokai Kosho Co., Ltd.) containing less than 1000 ppm nitrogen impurities was used as the carbon source for diamond synthesis.
After crystal growth, the sample was cut and polished to obtain a (100)-oriented substrate. 
The mean concentration of substituted nitrogen (P1 centers), evaluated by electron spin resonance spectroscopy (JES-FA100, JEOL Co., Ltd.)~\cite{ESR}, was 8 ppm. 
Fig.~\ref{fig:DiamondPhoto} (a) shows a photograph of the diamond substrate, which is approximately 2.5 $\times$ 2 mm in size and 0.20 mm in thickness.
The black spots in the photograph represent metallic inclusions formed during the diamond synthesis process.
The yellow regions indicate areas of high nitrogen concentration.
Alpha-ray imaging was performed by selecting regions with strong luminescence and avoiding areas with inclusions.
The radio-luminescence spectrum measured using an X-ray source and the transmittance spectrum are shown in Fig.~\ref{fig:DiamondPhoto} (b). The details of each measurement setup are described in~\cite{DiamondScintillator1} and~\cite{XRadiMes}. These results represent average values within a central region of $\phi$ = 1 mm, and spatial variations are expected in the present sample.
From the luminescence spectrum, the color center was estimated to be associated with the Band A~\cite{BandA}.
The luminescence decay time constant was found to be on the millisecond scale. Although the long decay time complicates accurate evaluation of the light yield, it is expected to be comparable to, or even exceed, our previous result (5.5 $\times$ 10$^{4}$ photons/MeV),  based on a comparison of several luminescence measurements.

\subsection{microscopy imaging system}

\begin{figure}[htbp]
 \begin{center}
 \includegraphics[width=10cm]{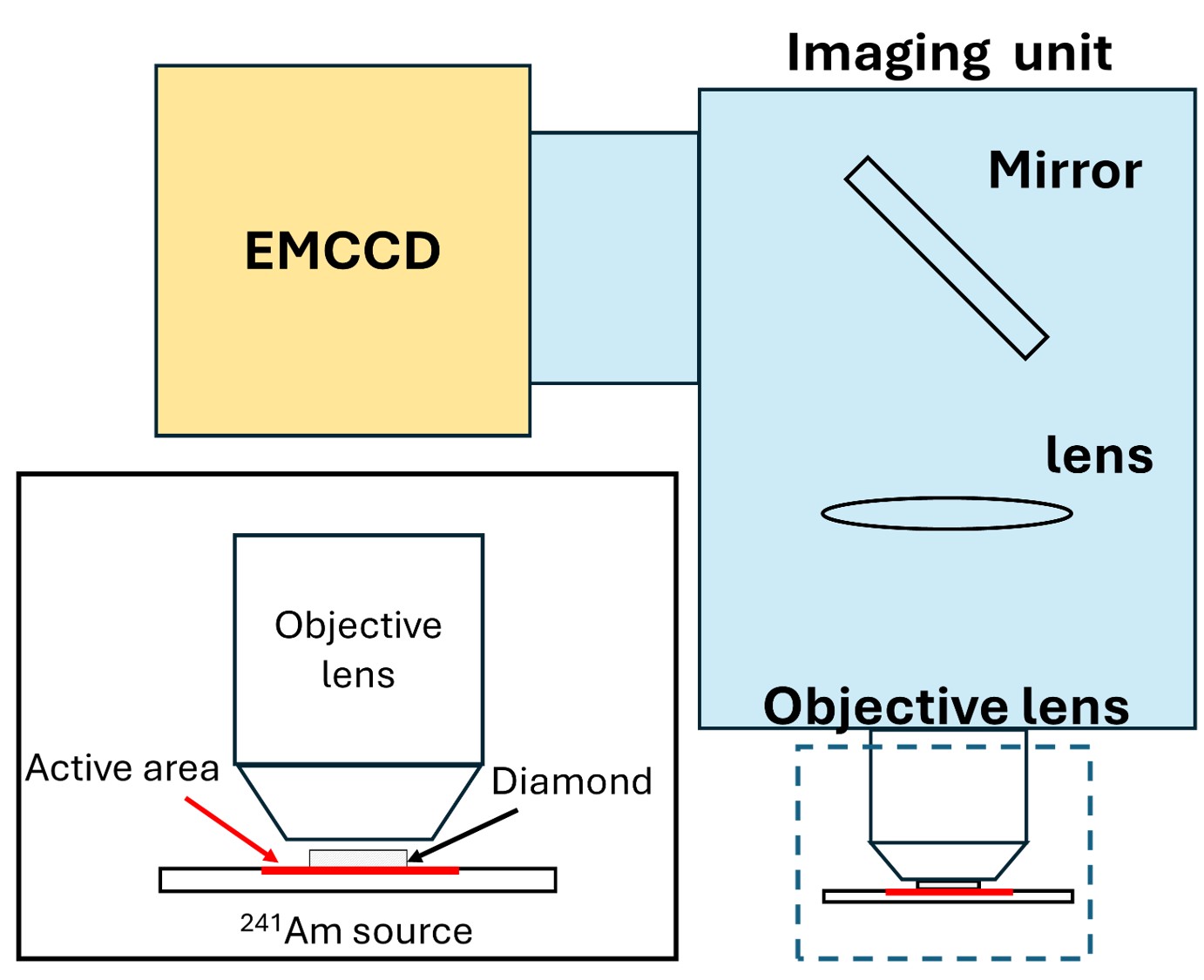}
\caption{Schematic drawing of developed alpha-ray imaging system.}
\label{fig:ImagingSystem}
 \end{center}
\end{figure}

The design of the imaging system was based on~\cite{TAPT_Yamamoto}. 
A schematic drawing is shown in Fig.~\ref{fig:ImagingSystem}.
The system was developed using commercially available equipment: an objective lens (CFI Plan Apo Lambda 40X, Nikon Corporation, Japan), an imaging unit (AA51, Hamamatsu Photonics, Japan), and an EMCCD camera (ImagEM C9100, Hamamatsu Photonics, Japan). We used a sealed $^{241}$Am source (AM162CE, Japan Radioisotope Association) with an activity of 3 MBq $\pm$ 20$\%$.
This source intensity ensures a sufficient number of events within the imaging field of view, making it easier to achieve proper focusing.
The diamond was placed directly above the active area of the alpha-ray source, as shown in the magnified view enclosed by the rectangle in Fig~\ref{fig:ImagingSystem}. Luminescence from the diamond was collected by the objective lens and imaged using the EMCCD camera.
The height of the objective lens (defined as the depth direction) could be adjusted using a stage controller, and imaging was performed at the stable depth position where the luminescence appeared brightest.
The measurements were performed under ambient conditions at room temperature and atmospheric pressure, with the entire system covered by a blackout curtain to block external light.
The numerical aperture (NA) of the objective lens was 0.95. Since no oil or grease was used, the focusing efficiency is reduced due to the refractive index of diamond ($n$ = 2.43 at $\lambda$ = 500 nm ~\cite{Zaitsev}).
A 16-bit monochrome image with a resolution of 512 $\times$ 512 pixels and a pixel size of 16 $\times$ 16 $\mu$m is obtained using the EMCCD.
This setup enabled the acquisition of optical images with a field of view of 204.8 $\times$ 204.8 $\mu$m, a digitization step of 400 nm.

\section{Imaging experiment}
\begin{figure}[htbp]
 \begin{center}
 \includegraphics[width=16cm]{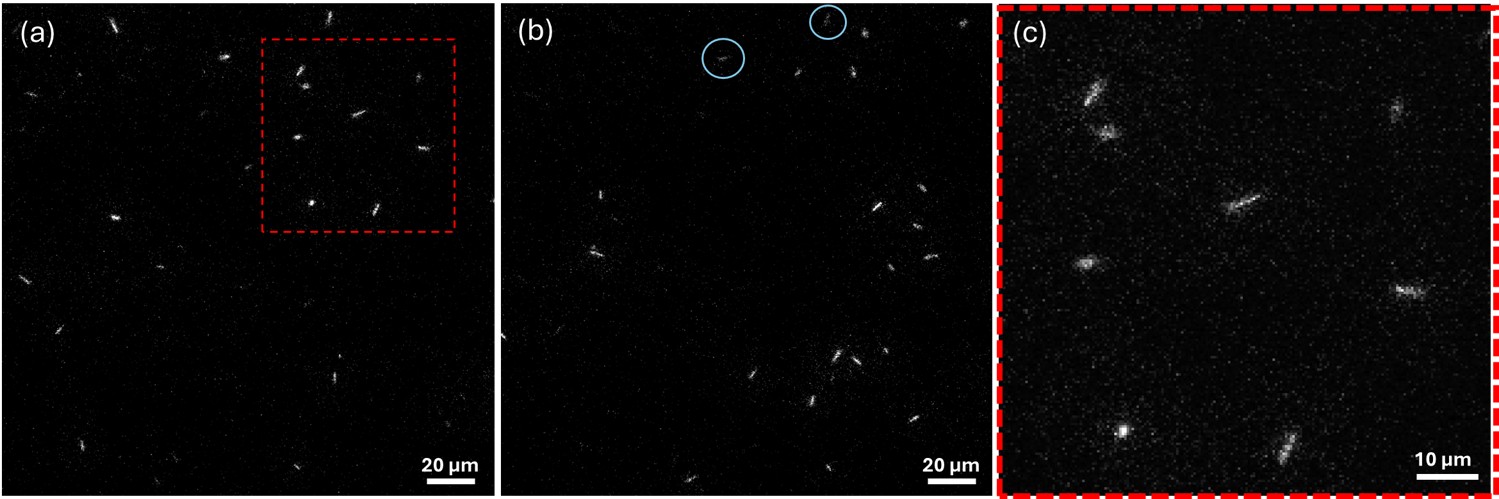}
\caption{Examples of detected images of alpha-ray tracks obtained via the luminescence from the diamond. (a) and (b) were acquired sequentially, and (c) shows a magnified view of the region enclosed by the red dashed line in (a). The circled events in (b) are identified as residual signals, as similar tracks were observed at the same positions in (a). }
\label{fig:exampleTrack}
 \end{center}
\end{figure}

Fig.~\ref{fig:exampleTrack} presents examples of raw images of alpha-ray tracks obtained in this experiment. 
The camera exposure time was set to 31 ms, and Fig.~\ref{fig:exampleTrack} (a) and Fig.~\ref{fig:exampleTrack} (b) were captured sequentially, with Fig.~\ref{fig:exampleTrack} (b) taken approximately 0.4 ms after Fig.~\ref{fig:exampleTrack} (a). Fig.~\ref{fig:exampleTrack} (c) shows a magnified view of the region enclosed by the red dashed line in Fig.~\ref{fig:exampleTrack} (a).
Although the image quality is limited, the alpha-ray tracks can be identified in the optical image.
The circled events in Fig.~\ref{fig:exampleTrack} (b)  may represent residual signals, as similar tracks with higher brightness were observed at the same positions in Fig.~\ref{fig:exampleTrack} (a). This is likely due to the millisecond-scale luminescence decay time of the diamond, which is comparable to both the camera exposure time and the interval between image acquisitions.

To enable automatic event recognition and track length detection, image processing methods were applied.
Given that the number of pixels in the detected tracks is comparable to those reported in~\cite{NIT_elli}, a previously developed method for analyzing super-fine grained nuclear emulsion~\cite{NIT} was employed with slight modification.
The image processing was conducted in two stages. In the first stage, clustering was applied to detect track-like optical signals, from which the spatial position and brightness values of each cluster were extracted. Signal and noise (residual signals) discrimination was then performed based on the mean brightness. 
In the second stage, frequency-domain filtering based on the discrete Fourier transform (DFT) was applied to enhance optical image clarity for track length detection, enabling isotropic processing regardless of the track angle ~\cite{NIT_elli}. This analysis requires cropped images of 51 $\times$ 51 pixels; therefore, events located within a 30-pixel margin from the edges of the raw images were excluded to ensure reliable processing. Additionally, events were excluded from the analysis if another track was located within a 12-pixel radius. The processing scheme of each method is described below.

\begin{figure}[htbp]
 \begin{center}
 \includegraphics[width=16cm]{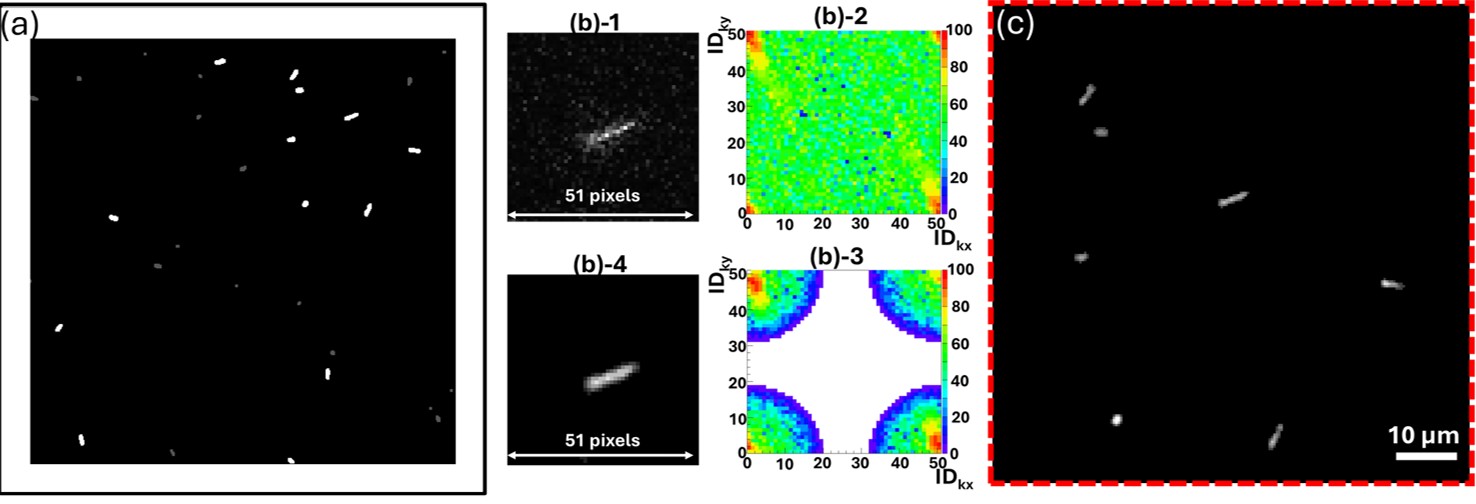}
\caption{Examples of processed images corresponding to Fig.~\ref{fig:exampleTrack}(a). (a) Output image of binarized
clusters obtained from the first-stage image processing. The color reflect the signal classification based on mean brightness: white denotes signal events, while gray indicates noise. (b) Result from the second-stage processing: (b)-1 raw cropped image (51 $\times $51 pixels) corresponding to an event located near the center of Fig.~\ref{fig:exampleTrack} (c); (b)-2  frequency power spectrum obtained via DFT, (b)-3 power spectrum after filtering, and (b)-4 final image after processing. (c) Reconstructed image displayed in the same view as Fig.~\ref{fig:exampleTrack} (c). }
\label{fig:exampleTrack2}
 \end{center}
\end{figure}

In the first stage, a Gaussian blur filter with a kernel size of 9 $\times$ 9 was first applied for smoothing. High-frequency components were then extracted by subtracting images before and after a low-pass filter. The low-pass filter, with a cutoff frequency of 0.08 cycles/pixel and Hamming window function of 1 $\times$ 15 and 15 $\times$ 1, was applied in both X and Y dimension. Subsequently, the binarization was performed by selecting pixels with brightness values greater than 30. The expansion after shrinking the image (morphological opening process) was then applied to eliminate residual spurious noise. 
For each binarized cluster with more than 4 pixels, the mean brightness of the corresponding area in the raw image was obtained.

\begin{figure}[htbp]
 \begin{center}
 \includegraphics[width=10cm]{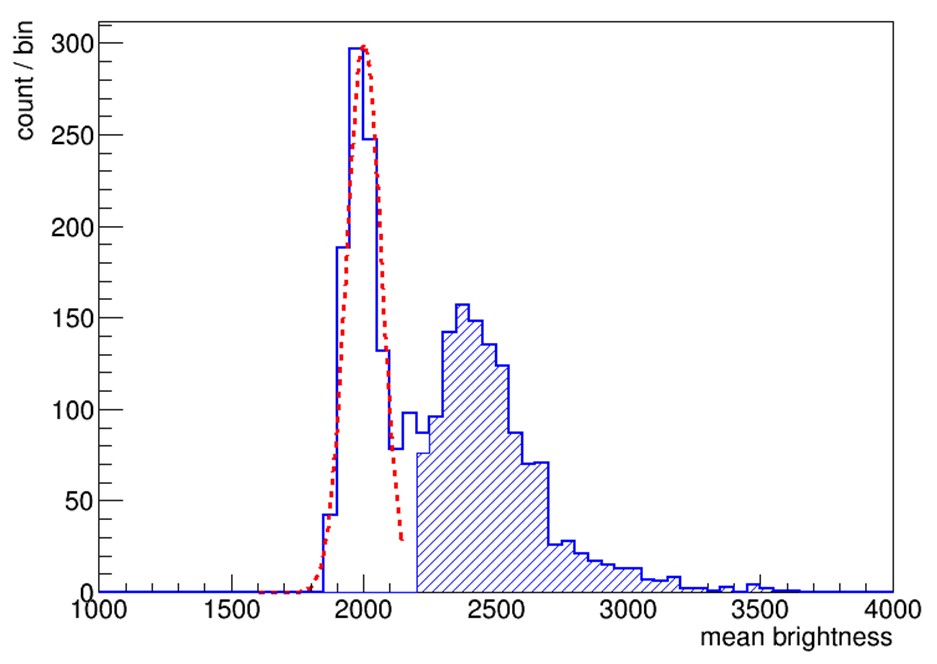}
\caption{Mean brightness distribution of optical image of detected events. The noise region was determined from the lower brightness part of the distribution, and the diagonally shaded area represents the signal region.}
\label{fig:brightness}
 \end{center}
\end{figure}

Fig.~\ref{fig:brightness} shows the distribution of mean brightness values for detected events across 100 sequential images.
The distribution appears to consist of two components: signal and noise. 
While the lower brightness region is assumed to correspond to noise, it may also include low-brightness signal events. 
To define the noise region, a Gaussian function was fitted to the distribution. The red dashed line in Fig.~\ref{fig:brightness} represents the fitting result.
The fitted mean and sigma were obtained as 2003 and 68, respectively. Using these values, the noise region was defined as the range within three sigma of the mean.
Consequently, events with a mean brightness greater than or equal to 2207 were selected as signal events, as indicated by the diagonally shaded region in Fig.~\ref{fig:brightness}. Fig.~\ref{fig:exampleTrack2} (a) presents the binarized cluster image obtained from the first-stage image processing. All events that were visually identified in Fig.~\ref{fig:exampleTrack} (a) were correctly recognized. The color reflect the signal classification based on mean brightness: white denotes signal events, while gray indicates noise. Additionally, a 30-pixel-wide margin around the image edges was excluded from the analysis region and is shown in solid white.
After this initial selection, 1277 out of 2370 events remained for the next analysis.

\begin{figure}[htbp]
 \begin{center}
 \includegraphics[width=15cm]{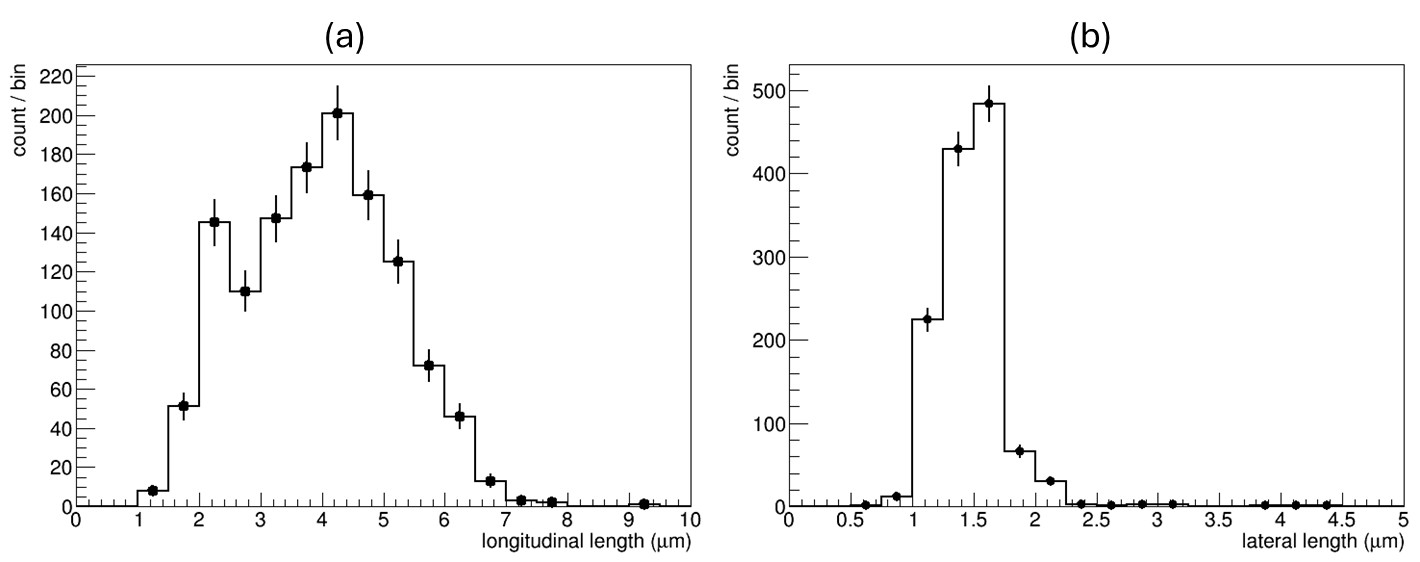}
\caption{Two-dimensional track lengths obtained from the optical images via image processing: (a) longitudinal length distribution and (b) lateral length distribution.}
\label{fig:track_data}
 \end{center}
\end{figure}

As an example, the result of the second-stage image processing is described for an event located near the center of Fig.~\ref{fig:exampleTrack} (c), which also appears in Fig.~\ref{fig:exampleTrack2} (b). The raw image of the selected event was cropped to a size of 51 $\times$ 51 pixels, as shown in Fig.~\ref{fig:exampleTrack2} (b)-1. This cropped image was then converted to the frequency domain using a DFT. Fig.~\ref{fig:exampleTrack2} (b)-2 shows the resulting frequency power spectrum, where low-frequency components appear in the four corners. The X and Y axes represent the wave number IDs, with a sign change occurring at ID$_{kx}$, ID$_{ky}$ = 25.
Frequency components correlated with the track shape are observed near (ID$_{kx}$, ID$_{ky}$) = (50, 0) and (0, 50). A filter optimized to extract these components, as described in ~\cite{NIT_elli}, was applied. 
The filtered power spectrum is presented in Fig.~\ref{fig:exampleTrack2} (b)-3.
An inverse DFT was then performed to reconstruct the spatial domain image, retaining only the components with brightness values greater than 2 for noise reduction. The resulting processed image is shown in Fig.~\ref{fig:exampleTrack2} (b)-4, which exhibits significantly improved clarity compared to the unprocessed image.
Fig.~\ref{fig:exampleTrack2} (c) displays the reconstructed image after the second-stage image processing, which also demonstrates improved clarity compared to Fig.~\ref{fig:exampleTrack} (c).
Following image processing, the contour with the largest area was extracted for each event. The height of the minimum-area enclosing rotated rectangle was defined as the longitudinal track length in the two-dimensional plane. 
For instance, the event in Fig.~\ref{fig:exampleTrack2} (b)-4 has a longitudinal length of 14.9 pixels (5.9 $\mu$m). 
The lateral range, defined as the width of the same rotated rectangle orthogonal to the track direction, was measured to be 3.7 pixels (1.5 $\mu$m).
Distributions of the two-dimensional track lengths are shown in Fig.~\ref{fig:track_data}:  Fig.~\ref{fig:track_data}(a) presents the longitudinal length distribution, while Fig.~\ref{fig:track_data}(b) shows the lateral range distribution.
The alpha-ray tracks exhibit high linearity, with lateral spreads estimated to be less than 200 nm.
Thus, the distribution in Fig.~\ref{fig:track_data}(b) primarily reflects the spatial resolution of the optical system, which is approximately 1.5 $\mu$m.
To summarize the number of detected events related to recognition efficiency, 1257 events were detected per 0.10 mm$^{2}$$\cdot$s of effective exposure, accounting for both the effective camera frame size and exposure time.

\section{Discussion}

\subsection{longitudinal track length and recognition efficiency}
To understand our experimental results, Monte Carlo simulations were conducted using Geant4 with the EmLowEnergy physics list.
The simulation geometry replicated the experimental setup illustrated in Fig.~\ref{fig:ImagingSystem}, where the diamond sample was positioned directly on top of the $^{241}$Am alpha-ray source.
The radiation source was sealed, and attenuation of alpha particles was expected due to the presence of a protective surface layer.
The composition of this protective layer was estimated based on information provided by the Japan Radioisotope Association and inferred from alpha spectrometry measurement using an ion-implanted silicon charged-particle detector.
The source structure was modeled as a layered configuration, from the surface inward, consisting of: (1) Au/Pa (1.5 $\mu$m), (2) Am oxide (2 $\mu$m), (3) Au/Pa (20 $\mu$m), and (4) Ag substrate (275 $\mu$m).
Taking this structure into account, the energy peak of the emitted alpha-rays was estimated to be approximately 4.5 MeV, consistent with both the alpha spectrometer measurements and the Geant4 simulation results.

\begin{figure}[htbp]
 \begin{center}
 \includegraphics[width=15cm]{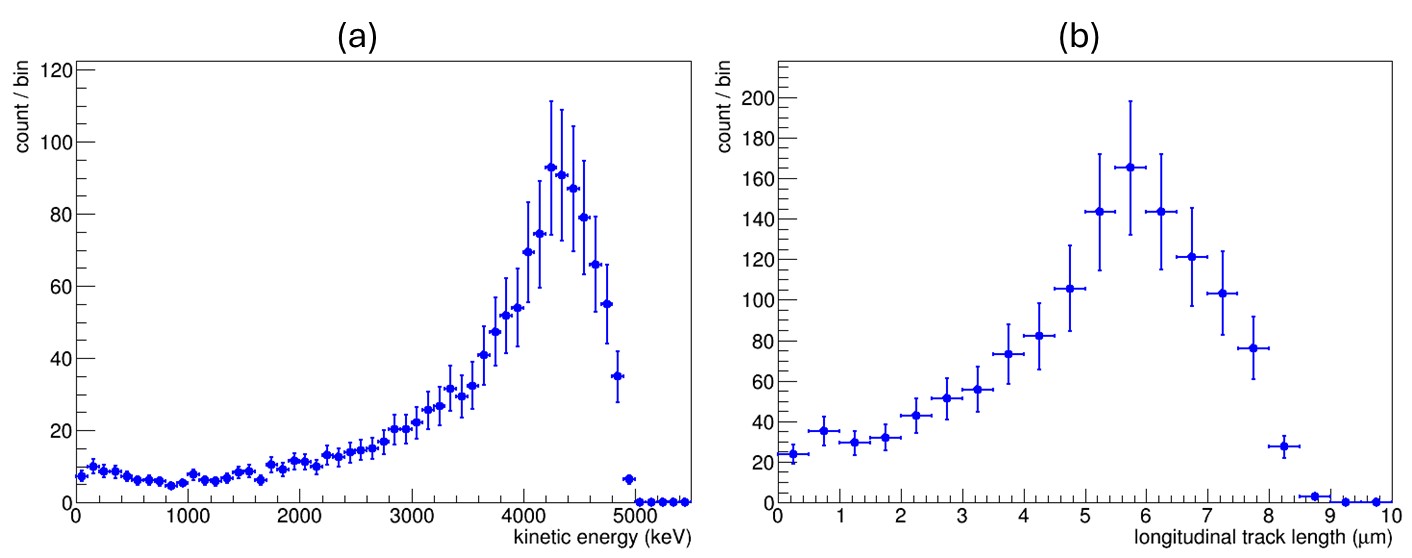}
\caption{Geant4 simulation results of alpha-ray tracks in diamond: (a) kinetic energy distribution of alpha-ray at the point of incidence on the diamond surface, and (b) distribution of longitudinal track lengths within the diamond.}
\label{fig:Geant4_1}
 \end{center}
\end{figure}

Fig.~\ref{fig:Geant4_1} shows the simulation outputs: (a) the kinetic energy distribution of alpha-rays upon reaching the diamond surface, and (b) the distribution of longitudinal track ranges within the diamond.
A uniform spatial distribution of alpha emission was assumed for the source. The overall effective exposure in the simulation was scaled to match the experimental data.
The simulation accounted for a 20$\%$ uncertainty in the alpha source activity, as specified by the Japan Radioisotope Association.

A comparison between the simulated and experimental longitudinal track length distributions shows that the experimental tracks are generally shifted toward shorter values, with the peak position difference of approximately 2 $\mu$m.
The current simulation does not yet account for the effects of optical resolution or depth of field (DOF); these effects will be incorporated in subsequent analysis.
The optical spatial resolution was assumed to be 1.5 $\mu$m, based on the experimental result shown in Fig.~\ref{fig:track_data}(b).
The DOF was estimated using the following equation~\cite{DOF}:

\begin{equation}
 DOF = \frac{ \lambda \cdot n}{NA^{2}} + \frac{n \cdot e}{NA}
\label{eq:2_n}
\end{equation}

where $e$ represents the digitization step of the imaging system.
As shown in Section 2.2, with NA = 0.95, $\lambda$ = 500 nm, $n$ = 2.43 and $e$ = 400 nm,  the DOF was calculated to be approximately 2.5 $\mu$m. It should be noted that these values are only rough approximations and may vary depending on the experimental conditions, as multiple interrelated factors affect both resolution and DOF.

\begin{figure}[htbp]
 \begin{center}
 \includegraphics[width=15cm]{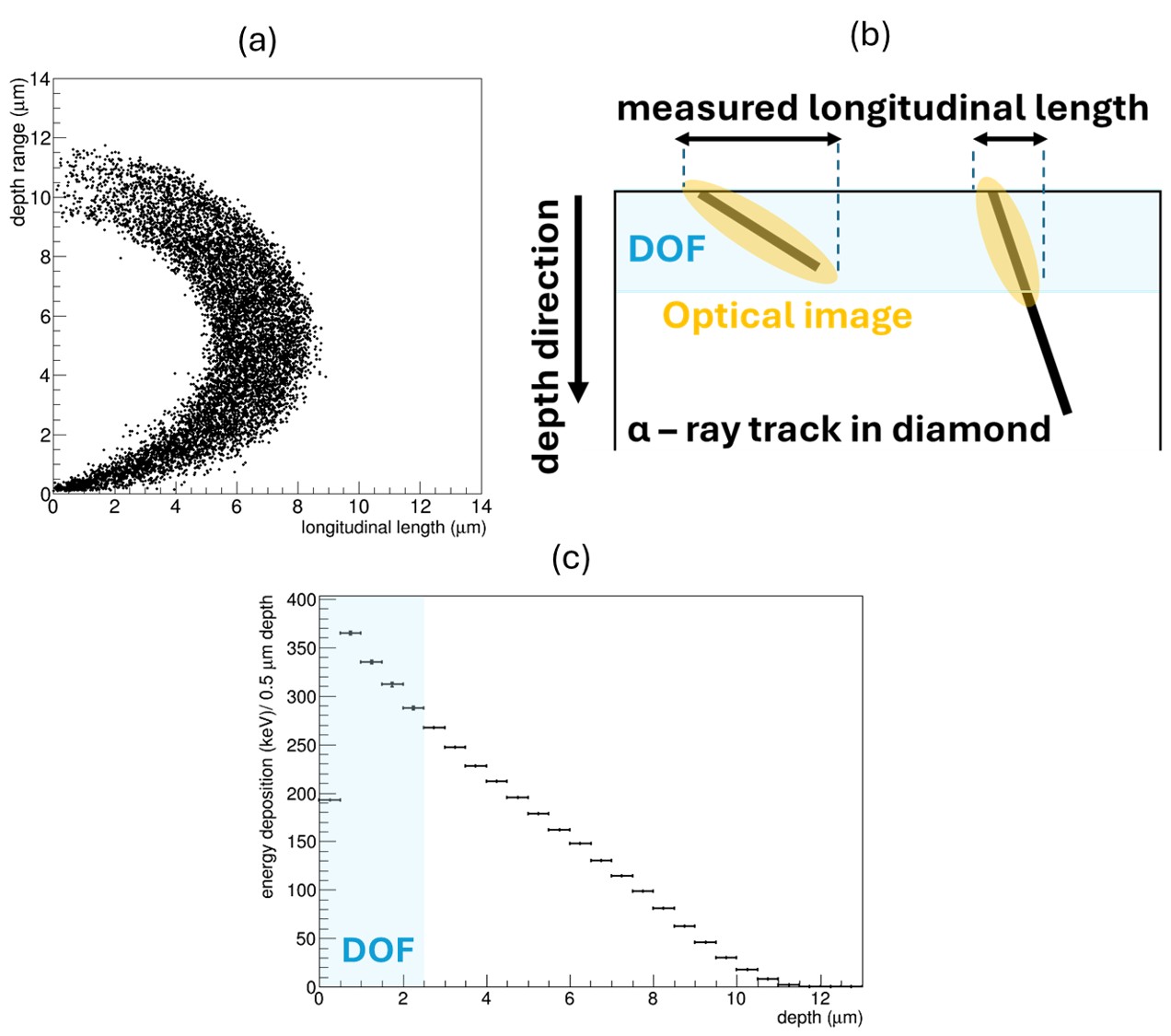}
\caption{Geant4 simulation results: (a)  relationship between the longitudinal and depth ranges, (b) schematic illustration of the concept of the measured longitudinal track length from optical imaging,  (c)  average energy loss distribution as a function of depth in a single alpha-ray track.}
\label{fig:Geant4_2}
 \end{center}
\end{figure}

Fig.~\ref{fig:Geant4_2} (a) shows the simulation result for the relationship between the longitudinal and depth ranges of alpha-rays tracks.
Owing to the source geometry, strictly horizontal tracks are rare, whereas those traveling along the depth direction can reach lengths of up to 12 $\mu$m.
In the experimental analysis, it was assumed that only the horizontal track components lying within the DOF at a specific depth were observable, appearing as optical images broadened by the optical resolution,  as illustrated in Fig.~\ref{fig:Geant4_2} (b).
This depth corresponds to the region with the strongest luminescence, where energy deposition is highest. 
Fig.~\ref{fig:Geant4_2} (c) shows the average energy deposition as a function of depth, calculated for each individual track.
The horizontal axis represents the depth from the diamond surface (i.e., the side facing the radiation source), while the vertical axis indicates the energy deposited per 0.5 $\mu$m depth segment in a single alpha-ray track.
The results reveal higher energy loss near the surface, suggesting that the focal plane of the microscope was located in this region. 
Consequently, the DOF region was defined as 2.5 $\mu$m from the surface, and the longitudinal track lengths within the DOF were extracted from the simulation.

\begin{figure}[htbp]
 \begin{center}
 \includegraphics[width=11cm]{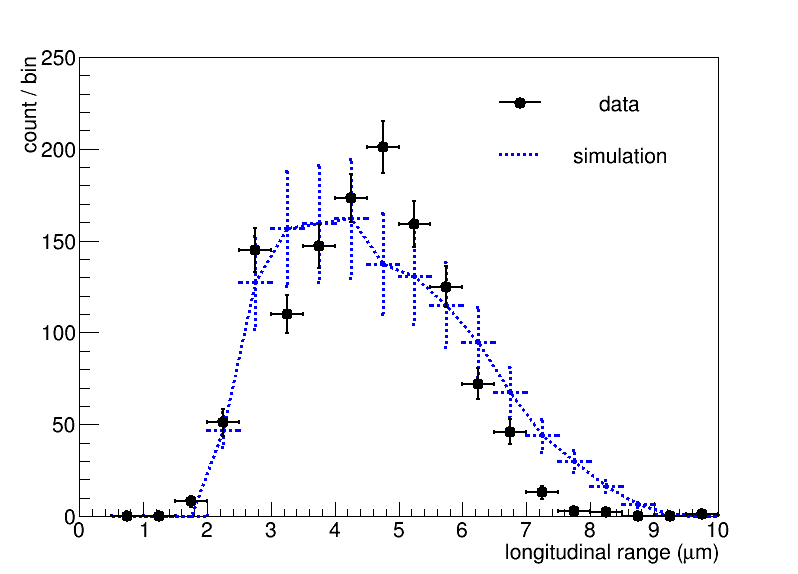}
\caption{Comparison of longitudinal track range distributions between experimental data and simulation with optical diffraction taken into account.}
\label{fig:Geant4_3}
 \end{center}
\end{figure}

Fig.~\ref{fig:Geant4_3} compares the experimental and simulated longitudinal track length distributions. 
In the simulation, the track lengths were adjusted by adding the 1.5 $\mu$m optical resolution to account for spatial broadening due to the imaging system.
Although a more detailed quantitative evaluation of track appearance under the given optical conditions is required, the overall shape of the experimental track length distribution is broadly consistent with the simulation. 
These results indicate that the current imaging system offers sufficient performance for capturing and visualizing alpha-ray tracks in real time.
The recognition efficiency, determined by comparing the observed number of events with the expected number from the simulation, was 97$\pm$19$\%$. 
Although uncertainties remain in the source intensity and noise rejection (contamination) rate, the present results constitute a sufficient first demonstration toward realizing high‑efficiency alpha-ray imaging.

\newpage

\subsection{Future prospects}
In previous studies, ZnS(Ag) scintillators demonstrated point‑like imaging with a x5 objective lens, achieving a spatial resolution of 16.2 $\mu$m ~\cite{alpha_kurosawa}. In contrast, our system resolves alpha-ray tracks with an optical broadening of 1.5 $\mu$m, representing an order‑of‑magnitude improvement in resolution.
Compared to GAGG, there is no significant difference in the range distribution or image appearance, indicating that diamond exhibits performance comparable to that of GAGG. Moreover, diamond offers superior radiation hardness, making it promising for applications that exploit this property.
Previous studies on alpha-ray imaging did not implement automatic track length detection. The automatic image processing method developed in this study could thus significantly advance research in this area. Our method enables event selection and track length extraction in less than one second per field of view, allowing not only real-time image acquisition but also simultaneous detection of track information. 
To determine the reaction point, intentionally shifting the focus toward the surface enables identification of the start and end points based on the broadening of the optical image ~\cite{JINST_Yamamoto}. This approach allows for reaction point detection with an accuracy of approximately 1.5 $\mu$m, which is comparable to the optical resolution.
Furthermore, employing an alternative optical analysis method—such as peak position determination based on the brightness distribution—can enhance the position determination accuracy to below the submicron level, even without the use of a higher magnification lens.

As for improvements in diamond devices, developing uniform crystals without inclusions would enable consistent alpha-ray detection regardless of location within the crystal, facilitating broader adoption of this technology by general users. Improvement in luminescence decay time is also desirable; for applications involving higher radiation dose rates, a decay time in the nanosecond to microsecond range would be more suitable than the current millisecond-scale response. It has been found that the light yield and decay time of diamond scintillators vary depending on impurity content. A detailed report on the correlation between nitrogen impurities and scintillation properties is scheduled for publication in the near future.

\section{Conclusion}
We have demonstrated real-time alpha-ray track imaging using a diamond scintillator plate synthesized in our laboratory. Alpha-ray tracks were imaged from the luminescence of the diamond using a x40 objective lens and an EMCCD under ambient conditions (room temperature and atmospheric pressure). We successfully captured alpha-ray track images with an optical resolution of 1.5 $\mu$m. The two-dimensional track length was subsequently measured via image processing and compared with a Geant4-based simulation that accounted for the effects of optical diffraction. The overall shape of the track length distribution in the micrometer range was found to be in general agreement with the simulation, and a recognition efficiency approaching 100$\%$ was achieved. These results indicate that the current imaging system provides more than sufficient performance for real-time detection and visualization of alpha-ray tracks, highlighting the potential of diamond as a scintillator material for advanced radiation imaging applications.

\section*{Acknowledgement}
We thank Takuya Shiraishi (Kanagawa University Faculty of Science) for his help with the Geant4 simulations.
We also thank Chikara Shinei (University of Tsukuba)  for his support with the ESR measurements.
This work was supported by Japan Society for the Promotion of Science (JSPS) KAKENHI Grant-in-Aid for Early-Career Scientists 24K17075. This work was performed at Research Divisions and Groups under the GIMRT Program of the Institute for Materials Research, Tohoku University (Proposal No. 202412-RDKYA-0061).

\end{document}